\documentclass[11pt,letterpaper]{article}
\usepackage[utf8]{inputenc}
\usepackage{amsmath}
\usepackage{amsfonts}
\usepackage{amssymb}
\usepackage{nuprl}
\title{Formalizing Category Theory and Presheaf Models of Type Theory in Nuprl}
\author{Mark Bickford, Cornell University Computer Science}

\begin{document}
\maketitle
\begin{abstract}
This article is the first in a series of articles that explain the
formalization of a constructive model of \emph{cubical type theory} 
in Nuprl\footnote{This work supported by NSF EAGER Award 1650069: Constructive Univalent Foundations}.
In this document we discuss only the parts of the formalization that
do not depend on the choice of base category. So, it spells out
how we make the first steps of our formalization of cubical type 
theory.
\end{abstract}
\section{Introduction}
Vladimir Voevodsky's addition of \emph{univalence} to type theory
has generated much interest. For an overview, one can consult
the HoTT book \cite{HoTT}.
A constructive model of a type theory satisfying univalence 
was first created by Bezem,Coquand, and Huber \cite{BCH}. It is
a \emph{presheaf model} over a category of cubical sets.
We spent some time formalizing this, the BCH model, in Nuprl \cite{book-full}\footnote{details at http://www.nuprl.org/wip/Mathematics/cubical!sets/index.html} but
were not able to complete the formal verification of the
crucial composition property for the universe (which took over twenty pages in Huber's thesis).

Cohen, Coquand, Huber, and M\"ortberg created \emph{cubical type theory} \cite{cubicalTT} and a semantics for it that is a
presheaf model using a new base category of cubes based on free deMorgan algebras. With NSF support, we were able to formalize this
model in Nuprl and prove that all of its rules are correct, and
have constructive content.
This article is the first in a series of articles that explain that
formalization effort, which includes the longest proofs ever done in Nuprl.

\section{Overview}
The main features of cubical type theory that will
concern us at first are that in addition to the {\tt \mPi} and {\tt \mSigma} types of dependent functions and dependent pairs, there will be
a new type {\tt \mBbbI} for the \emph{interval}. When there are
several variables {\tt i,j,k:\mBbbI} in the context, then those parameters range over (n-dimensional) \emph{cubes}. The theory will
also have \emph{formulas} about these variables like {\tt (i=0 and j=1)} or more complex formulas including clauses like {\tt k = min(i,j)}. If we call the 
whole n-dimensional cube {\tt X} and the subspace defined by the formula {\tt S}, then if we add a new  variable {\tt v:\mBbbI}, we get spaces {\tt X+ = X \mtimes{} \mBbbI} and {\tt S+ = S \mtimes{} \mBbbI}, and 
the subspace of {\tt X+} for {\tt v=0} is a copy of {\tt X} 
we can call {\tt X[0]}. If the subspace {\tt S+ \mcup{} X[0]} 
is a \emph{homotopy retract} of the whole cube {\tt X+} then
any continuous function defined on {\tt S+ \mcup{} X[0]} can be extended to the whole space. Daniel Kan repeatedly used this \emph{homotopy extension principle} in his development of homotopy
theory.  The principle is expressed 
synthetically in the rules of cubical type theory and to give a
constructive model of cubical type theory we must produce the
witnesses (i.e. the programs) for these extension operations.
This is accomplished by endowing each type in the theory with a \emph{composition operation} (from which the extension operation can
be derived). Types with a composition operation are called \emph{fibrant} (or said to satisfy the \emph{Kan property}). When all the
types in the \emph{universe} are fibrant and the universe is itself fibrant then it will be possible to verify Voevodsky's \emph{univalence axiom}. Showing that all the types are fibrant (especially the universe) is the most technically demanding part of
construction of a constructive model of cubical type theory and is  the topic of later articles in this series.

Eventually we will define
the \emph{paths} in a type {\tt T} to be the functions of type {\tt \mBbbI \mrightarrow{} T}, and a path with \emph{endpoints} {\tt a} and {\tt b} will witness equality of {\tt a} and {\tt b}. This will give
a new meaning for equality that has non-trivial computational content.
It differs from the meaning of equality in Nuprl, and a constructive
model of this theory is highly non-trivial. 

The model we formalized is due to Bezem, Coquand, Cohen and M\"ortberg
but starts with a construction due to Hofmann and Streicher, the \emph{presheaf models} of Martin L\"of type theory. The point of the
current article is to explain this construction and how it is formalized in Nuprl. A by-product of this explanation will be a sort of
\emph{Rosetta stone} relating the category-theoretic viewpoint and language to the purely type-theoretic viewpoint and language used in
Nuprl. So, for example, a \emph{presheaf} is seen to be the same as
a \emph{family of Nuprl types} together with a family of maps between
them that satisfy certain equations. The category-theoretic language can say a lot with one word like
\emph{presheaf} while the type theoretic definitions are concrete
(and understood by Nuprl tactics!). Those who already know the category theory can see how it is expressed in Nuprl, while those
who know a type theory like Nuprl's can learn what the category theory
means.

Formalizing category theory in Nuprl is fairly straightforward.
Each category will be small---in the sense that it is a type in
some Nuprl universe. But since Nuprl has an infinite cumulative hierarchy of universes that does not seem to limit the theory.

We have formalized basic concepts of category theory sufficient
for our construction of a formal model of cubical type theory.
For this we mainly needed the concept of a \emph{presheaf}.
So we need categories, functors, natural transformations,
the opposite of a category, and a ``category of sets''.
For the latter, we substitute the category of Nuprl types 
(in a universe). Because Nuprl's types are extensional, the definition of presheaf that we get has the properties needed
to build constructive models of type theory and to prove things
like the Yoneda lemma---that
the Yoneda embedding is a full-faithful functor.
We have also defined adjoint functors, monads, Kleisli category,
groupoids, comma categories and many other concepts, and constructed
examples such as the adjunction \verb%FreeGroup -| ForgetGroup%.

Once we have formalized basic category theory, the first step
in constructing a model of cubical type theory is to construct
a presheaf model of basic Martin L\"of type theory. This construction, due to Hofmann and Streicher, interprets the
basic concepts of MLTT: \emph{context, type, term} and its
judgments, using for contexts presheaves over any base category {\tt C}. This generic construction also defines the $\Pi$ and $\Sigma$ types and the basic terms for $\lambda$-abstraction and pairs with
first and second components (it also gives an interpretation of
equality types but for cubical type theory we will use  path types instead). 

The model for cubical type theory
starts with this basic construction and specializes it to use
a particular {\tt CubeCat} (the \emph{cube category}) for the
base category.
In this document we discuss only the parts of the formalization that
do not depend on the choice of base category. So, it spells out
how we make the first steps of our formalization of cubical type 
theory.

\section{Categories}

A category has a set or class of objects. For objects $x$ and $y$
there is a set or class of
arrows $x \rightarrow y$. For each object $x$ there is an identity
arrow $x \rightarrow x$, and arrows $x \rightarrow y$ and $y \rightarrow z$ can be composed to get an arrow $x \rightarrow z$.
We replace `set or class` by type (in a universe), so our 
formal definition in Nuprl has a universe \emph{level parameter} $i$.

There are several ways to build structures in Nuprl's type theory,
but to define categories we use the most straightforward one.
 We make a
dependent product for the objects, arrows, identity, and composition, and then, using Nuprl's refinement type (also called the \emph{set type}),
form a subtype of the dependent product for which certain
\emph{identity} and \emph{associativity} equations hold. We write
either {\tt Type\{i\}} or {\tt \mBbbU\{i\}} for the {\tt i}th universe.
\begin{program}
SmallCategory\{i\}  ==\\
  \{cat:ob:Type\{i\}\\
    \mtimes arrow:ob \mrightarrow ob \mrightarrow Type\{i\}\\
    \mtimes x:ob \mrightarrow arrow(x, x)\\
    \mtimes x:ob \mrightarrow y:ob \mrightarrow z:ob\\
       \mrightarrow arrow(x,y)\mrightarrow arrow(y,z)\mrightarrow arrow(x,z) |\\ 
    let ob,arrow,id,comp = cat\\ 
    in \mforall{}x,y:ob. \mforall{}f:arrow(x,y). comp(x,x,y,id(x),f) = f\\
                             \mwedge comp(x,y,y,f,id(y)) = f\\
    \mwedge \mforall{}x,y,z,w:ob. \mforall{}f:arrow(x,y) \mforall{}g:arrow(y,z). \mforall{}h:arrow(z,w).\\
          comp(x,z,w,comp(x,y,z,f,g),h) =\\
          comp(x,y,w,f,comp(y,z,w,g,h))\\
  \} 
\end{program}
The four components of a category {\tt C} are {\tt ob(C)}, {\tt arrow(C)} , {\tt id(C)}, and {\tt comp(C)}.
To make a category we supply its four components using a {\tt mk-cat} operator displayed as follows:
\begin{program}
Cat(ob               =  ob;\\
    arrow(x,y)       =  arrow[x;y];\\
    id(a)            =  id[a];\\
    comp(u,v,w,f,g)  =  comp[u;v;w;f;g])
\end{program}
The expressions on the right of the equal signs are second order
variables. The result is a category if the identity and associativity equations hold for the given expressions.
For example the \emph{discrete category} for a type {\tt X} is
\begin{program}
discrete-cat(X)  ==\\
Cat(ob= X; arrow(x,y)=  x=y; id(a)= \mcdot{}; comp(u,v,w,f,g)= \mcdot{})
\end{program}
This is the category with objects {\tt X} and arrows only between
equal members of {\tt X} (the equality type in Nuprl is inhabited
only by \mcdot{}, sometimes written {\tt Ax}).

A somewhat more interesting category is the category of types:
\begin{program}
TypeCat\{i\}  ==\\
Cat(ob= Type\{i\};\\
   arrow(I,J)=  (I \mrightarrow{} J);\\
   id(I)= \mlambda{}x.x;\\
   comp(I,J,K,f,g)= (g o f) )
\end{program}
This category has a universe level parameter {\tt i} and its objects
are the Nuprl types in universe {\tt i},
{\tt I \mrightarrow{} J} is the Nuprl function type, and
{\tt g o f} is {\tt \mlambda x.g(f(x))}. We use this category as
a replacement for the category of sets.

The \emph{opposite} of category {\tt C} is 
\begin{program}
OpCat(C)  ==\\
Cat(ob= ob(C);\\
   arrow(I,J)=  arrow(C)(J,I);\\
   id(I)= id(C)(I)\\
   comp(I,J,K,f,g)= comp(C)(K,J,I,g,f) )
\end{program}
It simply reverses the direction of all the arrows.

The category of groups is:
\begin{program}
Group  ==\\
Cat(ob= Group\{i\};\\
   arrow(G,H)=  MonHom(G,H);\\
   id(G)= \mlambda{}x. x;\\
   comp(I,J,K,f,g)= (g o f) )
\end{program}
{\tt MonHom(G,H)} is the type of \emph{monoid homomorphisms}, i.e.
maps that preserve the group identity and group operation.
\section{Functors}
A \emph{functor} {\tt F} between categories {\tt C1} and {\tt C2}
is a member of the following type:
\begin{program}
Functor(C1;C2)  ==\\
  \{FM:F:ob(C1) \mrightarrow ob(C2) \mtimes\\
       (x:ob(C1) \mrightarrow y:ob(C1) \mrightarrow \\
          (arrow(C1) x y) \mrightarrow (arrow(C2) (F x) (F y))) | \\
     let F,M = FM in\\ 
     \mforall{}x:ob(C1). M(x,x,id(C1)(x)) = id(C2)(F x)  \mwedge{}\\
     \mforall{}x,y,z:ob(C1). \mforall{}f:arrow(C1) x y.  \mforall{}g:arrow(C1) y z.\\
      M(x,z,comp(C1)(x,y,z,f,g)) =\\
      comp(C2)(F(x),F(y),F(z,M(x,y,f),M(y,z, g))\} 
\end{program}
{\tt F} has two components {\tt ob(F)} and {\tt arrow(F)}, where
{\tt ob(F)} maps objects of {\tt C1} to objects of {\tt C2},
and {\tt arrow(F)} maps arrows of {\tt C1} to arrows of {\tt C2}.
The functor must map identity arrows in {\tt C1} to identity arrows in {\tt C2} and map the composition of arrows in {\tt C1} to composition of arrows in {\tt C2}.
We display {\tt ob(F)(x)} as {\tt F(x)} and display {\tt arrow(F)(x,y,a)} as {\tt F(x,y,a)}.
To construct a functor we use {\tt mk-functor} which is displayed
\begin{program}
functor(ob(a)= ob[a];\\
       arrow(x,y,f)= arrow[x;y;f] )
\end{program} where the expressions on the right of the equal signs are second order
variables. For example, the identity functor (which we display as {\tt 1}) is {\tt functor(ob(x)=x; arrow(x,y,a)=a)}. Composition of
functors {\tt F} and {\tt G} is the functor
\begin{program}
functor(ob(x)= G(F(x)); arrow(x,y,a)= G(F(x),F(y),F(x,y,a)))
\end{program}
The identity functor and functor composition satisfy the equations
needed to define the \emph{category of categories}
\begin{program}
CatCat\{i\}  ==\\
Cat(ob= SmallCategory\{i\};\\
   arrow(A,B)=  Functor(A,B);\\
   id(A)= 1;\\
   comp(A,B,C,F,G)= functor-comp(F,G) )
\end{program}
This category is a member of the type {\tt SmallCategory\{i+1\}}.

A functor {\tt F \mmember{} Functor(C,D)} is \emph{full and faithful} if for any {\tt x,y \mmember ob(C)}, {\tt arrow(F)} is a bijection between {\tt arrow(C)(x,y)} and {\tt arrow(D)(F(x),F(y))}.
\section{Natural Transformations}
A \emph{natural transformation} between two functors {\tt F} and {\tt G} in {\tt Functor(C,D)} is a function that assigns to each
object {\tt A} in category {\tt C} an arrow in category {\tt D}
between {\tt F(A)} and {\tt G(A)} for which a certain diagram commutes (i.e. a certain \emph{naturality} equation holds). Thus a natural transformation is a member of the type:
\begin{program}
nat-trans(C;D;F;G)  ==\\
 \{trans:A:ob(C) \mrightarrow arrow(D)(F(A), G(A)) | \\
   \mforall{}A,B:ob(C). \mforall{}g:cat-arrow(C)(A,B).\\
   comp(D)(F(A),G(A),G(B),trans(A),G(A,B,g))\\
   = comp(D)(F(A),F(B),G(B),F(A,B,g),trans(B))\\
 \} 
\end{program}
A natural transformation {\tt T} is simply a function $\lambda x.T(x)$, but we use a special operator {\tt mk-nat-trans} displayed as 
{\tt x |\mrightarrow{} T[x]} to tell the system to type-check it
as a natural transformation. For example, the identity natural
transformation is
\begin{program}
identity-trans(C;D;F) ==  A |\mrightarrow{} id(D)(F(A))
\end{program}
Composition of natural transformation {\tt t1 \mmember{} nat-trans(C;D;F;G)} with natural transformation  {\tt t2 \mmember{} nat-trans(C;D;G;H)} is:
\begin{program}
t1 o t2 ==  A |\mrightarrow{} comp(D)(F(A), G(A), H(A), t1(A), t2(A))
\end{program}
This composition operator is really 
{\tt trans-comp(C;D;F;G;H;t1;t2)}, but we ``hide'' the parameters
{\tt C,D,F,G} and {\tt H} in the display form and display only
{\tt t1 o t2}.

The identity natural transformation and the composition operation
satisfy the equations needed to define the \emph{category of functors}
\begin{program}
FUN(C1;C2)  ==\\
 Cat(ob= Functor(C1;C2)\\
    arrow(F,G) = nat-trans(C1;C2;F;G)\\
    id(F) = identity-trans(C1;C2;F)\\
    comp(F,G,H,t1,t2)=  t1 o t2 )
\end{program}
\section{Presheaves}
A \emph{presheaf} over a category {\tt C} is a functor from
{\tt OpCat(C)} to the category of sets. We substitute the category
of types (in universe {\tt i}) for the category of sets. Thus
\begin{program}
Presheaf(C)\{i\} == Functor(op-cat(C);TypeCat\{i\})
\end{program}
This is a type in Nuprl universe {\tt i+1}. The presheaves over {\tt C} form a category
\begin{program}
Presheaves(C)\{i\} == FUN(op-cat(C);TypeCat\{i\})
\end{program}
This is a member of the type {\tt SmallCategory\{i+1\}}.

Since a presheaf is a functor, to construct one we must give the
two components. We write this
\begin{program}
Presheaf(Set(I) = S[I]\\
        Morphism(I,J,f,rho) = morph[I,J,f,rho]\\
        )
\end{program}
The expression {\tt S[I]} specifies the ``set'' (i.e. the type)
assigned to object {\tt I} from category {\tt C}. The expression
{\tt morph[I,J,f,rho]} specifies how to map the set {\tt S[I]} to 
the set {\tt S[J]} when there is an arrow {\tt f:J \mrightarrow{} I} in
category {\tt C} (the arrow is reversed because the presheaf is
a functor from the opposite of {\tt C}) by giving the image of {\tt rho \mmember{} S[I]} under the mapping. This map is called the \emph{restriction map} and we display it as simply {\tt f(rho)},
but it really has parameters {\tt H,I,J,f,rho} where {\tt H} is
the presheaf.

For example, the \emph{representable presheaf} for {\tt X \mmember{} ob(C)} is 
\begin{program}
Yoneda(X) == \\
Presheaf(Set(I) = arrow(C)(I,X)\\
        Morphism(I,J,f,a) = comp(C)(J,I,X,f,a)\\
        )
\end{program}
This presheaf assigns to each object {\tt I \mmember{} C} the set of
arrows {\tt I \mrightarrow{} X}. Given an arrow {\tt f:J \mrightarrow{} I}, composition with {\tt f} maps an arrow {\tt a \mmember{} I \mrightarrow{} X}
to an arrow {\tt b \mmember{} J \mrightarrow{} X}, and the 
equations necessary for this to define a presheaf hold.

We call this presheaf {\tt Yoneda(X)} because
it is the first component of the \emph{Yoneda embedding}:
\begin{program}
Functor(ob(X) = Yoneda(X)\\
        arrow(X,Y,f) = A |\mrightarrow \mlambda g. comp(C)(A,X,Y,g,f)\\
        )
\end{program}
For any category {\tt C}, this defines a functor from {\tt C} to
the category of presheaves over {\tt C}. The \emph{Yoneda lemma}
states that this functor is full and faithful. The proof of this
lemma in Nuprl was straightforward. It does make use of some of the
\emph{extensional} properties of Nuprl's type theory, which we will
discuss briefly in the next section.

\section{Extensional reasoning in Nuprl}
There are at least two different meanings for the adjective \emph{extensional} in type theories. One is \emph{function extensionality} and the other is sometimes phrased as ``\emph{propositional equality is definitional equality}''. We
explain how Nuprl's type theory is extensional in both of these senses. 

Every Nuprl proof is built up (by using \emph{tactics})
as a tree of \emph{primitive inferences} that are instances
of the \emph{rules}. A rule matches the current \emph{goal sequent}
with a given \emph{goal pattern} and then, using some (possibly empty) list of \emph{parameters} (supplied by the tactic), generates the instances of the \emph{subgoal patterns}.
The rule is true when the \emph{truth} of any instance of the
goal follows from the truth of the instances of the subgoals.
The formal definition of correctness for Nuprl rules thus depends
on the formal definition of \emph{truth} of a Nuprl sequent. This
in turn depends of the formal definition of the Nuprl type system.
All of this has been formalized in Coq and is well beyond the scope
of this document. Here we want to show the two rules that are
true for the Nuprl type system and show that it is extensional in both senses.

\paragraph{Function Extensionality}

\begin{program}
H \mvdash{} f = g \mmember (x:A \mrightarrow{} B)\\
\\
    BY  functionExtensionality  !parameter\{i:l\}  u \\
\\   
    H  u:A  \mvdash{}  f(u) = g(u) \mmember B[u/x]\\
    H \mvdash{}  A = A \mmember Type\{i\}
\end{program}
This rule says that to prove {\tt f} is equal to {\tt g} in
a (dependent) function type {\tt x:A \mrightarrow{} B} it is enough to prove that the
domain {\tt A} is a type and that for every {\tt u:A}, {\tt f(u)} and {\tt g(u)} are equal in {\tt B[u/x]} (the proof of that
subgoal will also establish that {\tt B[u/x]} is a type). This, the mathematical definition of function equality, is true
because of the formal definition of the (dependent) function type
in the Nuprl type system. Note that all items in the generated
subgoals come from matching the variables {\tt H, f, g, x, A}, and {\tt B}
with the the goal, except for the universe level {\tt i} and the
auxiliary variable {\tt u} that are supplied as parameters. In this
rule, all the judgments are equality types which have no constructive
content in Nuprl. Hence, no extract terms are specified (more precisely, the default extract {\tt Ax} is used).

\paragraph{Type Extensionality}
If types {\tt A} and {\tt B} are provably equal (i.e. \emph{propositionally equal}) and {\tt t \mmember{} A}
is it true that {\tt t \mmember{} B}? In \emph{intensional} type
theories this is usually not true unless {\tt A} and {\tt B} are
\emph{definitionally} equal. When that is not so, and {\tt p} is the proof of {\tt A = B}, some sort of
coercion function like {\tt transport(p,t) \mmember{} B} is
needed. In Nuprl we can prove that {\tt t \mmember{} B} without
applying any coercion.

This follows from the more general rule shown here:
\begin{program}
H  x:A, J  \mvdash{}  C  ext  t\\
\\
    BY  hyp\_replacement  \#j  B  !parameter\{i:l\}\\
\\   
    H  x:B, J  \mvdash{}  C  ext  t\\
    H  x:A, J  \mvdash{}  A = B \mmember Type\{i\}
\end{program}
In this rule, the parameter {\tt \#j} is the hypothesis number of
the declaration {\tt x:A} in the context. Because the conclusion {\tt C} may have constructive content, the \emph{extract term} {\tt t} is specified. In this case the rule says that the term extracted from the
proof of the original goal will be the term extracted from the proof
of the first subgoal.
The {\tt hyp\_replacement} rule says that in any context 
if type {\tt A} is provably equal to type {\tt B} (in some universe {\tt i}) then
we can replace a declaration {\tt x:A} in the context with
the alternate declaration {\tt x:B} (where {\tt B} is supplied as
a parameter). Not only is the original goal
true when the alternate goal is true, but the extract term {\tt t}
is the same because extracts are terms in
an untyped programming language.

\paragraph{Subtype reasoning}
Another distinctive feature of Nuprl's type theory is that
the \emph{subtype relation} {\tt A \msubseteq{} B} is defined by {\tt \mlambda{}x.x \mmember{} A \mrightarrow{}B} and is a proposition (i.e. a type) whenever {\tt A} and {\tt B} are types.\footnote{Membership {\tt t \mmember{} T} is just the
equality type {\tt t = t \mmember{} T}, but it is an interesting feature of Nuprl that sometimes (as when {\tt t} is {\tt \mlambda{}x.x}) an equality is well-formed whether it is true or not.}
The Nuprl library contains many lemmas about subtypes and reasoning
about subtypes is a signifcant part of the {\tt Auto} tactic.

The main way that type extensionality is used in Nuprl is via the
lemma {\tt subtype-rel-equal}:
\begin{program}
\mforall{}[A,B:Type]. A \msubseteq{} B  supposing A = B
\end{program}
The proof of this lemma uses the {\tt hyp\_replacement} rule.

For any lemma proved in Nuprl, the system can tell us which
lemmas and which
primitive rules were used in its proof. For the Yoneda lemma,
we found that the {\tt functionExtensionality} rule was used
and the {\tt subtype-rel-equal} lemma was used. 
So the Nuprl proof of the Yoneda lemma 
uses both kinds of extensional reasoning. 
We have not investigated whether both are absolutely
necessary for this proof because we are working in Nuprl, so
there is little point in knowing whether the natural proof of
a fact can be redone to avoid using some rule or other.

\section{Presheaf models of Martin L\"of Type Theory}
\label{basic}
Martin L\"of type theory (MLTT) has the following primitive concepts
expressed as \emph{judgements} of the formal theory.
\begin{itemize}
\item $(H \vdash)$ says that $H$ is a well-formed \emph{context}.
\item $(H \vdash T)$ says that $T$ is a well-formed \emph{type} in context $H$.
\item $(H \vdash t\!:\!T)$ says that $t$ is a well-formed  \emph{term} of type $T$ in context $H$.
\end{itemize}

A typical rule of MLTT would be that if $(H \vdash)$ and $(H \vdash T)$ then $(H, x\!:\!T \vdash)$ provided that $x$ is a fresh variable. This rule says that we can add new declarations to a
well-formed context, so starting with the \emph{empty context} (which is well-formed) contexts are built up as lists of declarations where
each type is well-formed in the preceding context. The well-formed
types and terms in such a context are certain syntactic expressions mentioning the declared variables.

To model such a theory exactly we would have to formally define
the syntax of the expressions and define free and bound variables,
$\alpha$-equality, and substitution. To avoid having to do this work, there is an alternate \emph{name-free} syntax for MLTT.
In this version rather than add $x\!:\!T$ to context $H$, we merely
add type $T$ to get the context $H.T$ and instead of the expression
$x$ in context $H, x\!:\!T$ a special term $q$ refers to the last declaration of the context $H.T$.
Rather than substitutions we use \emph{context maps} $\sigma \!:\!H \rightarrow G$. We can apply such a context map $\sigma$ to a type $T$ to get $(T)\sigma$ and to a term $t$ to get $(t)\sigma$.
If $G \vdash T$ then $H \vdash (T)\sigma$ and if $G \vdash t\!:\!T$ then $H \vdash (t)\sigma\!:\!(T)\sigma$.

There is a polymorphic context map $p\!:\!(H.T)\rightarrow H$. Thus, the variables $x,y,z$ in a context like $x\!:\!A,\ y\!:\!B,\ z\!:\!C$ correspond to the terms $((q)p)p,\ (q)p,\ q$ in the context
$A.B.C$. If we think of $q$ as zero and $p$ as successor, then
$((q)p)p$, $(q)p$, and $q$ correspond to the numbers $2, 1, 0$ which we
can see as \emph{DeBruijn indices} that replace the bound variables
$x,y,z$.

To build a model of this name-free version of MLTT we must
start by giving the meanings of $(H \vdash)$, $\sigma\!:\!H \rightarrow G$, $(H \vdash T)$, $(H \vdash t:T)$, $(H.T)$, $(T)\sigma$, $(t)\sigma$, $p$, and $q$. After that we define the function and product types (the $\Pi$ and $\Sigma$ types) and associated terms,
and prove that these definitions satisfy thirty-nine rules of basic MLTT that we give later (at the end of this section and in section~\ref{types}).

\paragraph{Context and context map} 
It was realized by Hofmann and Streicher that for any base category {\tt C} we can get a model
of MLTT where context $(H \vdash)$ means that $H$ is a presheaf
over {\tt C} and a context map $\sigma\!:\!H \rightarrow G$ is
a natural transformation from presheaf $H$ to presheaf $G$.

If we think about the Nuprl formalization of presheaf, we see that
a presheaf over category {\tt C} is a \emph{family of Nuprl types}
indexed by the objects in {\tt C} together with a family of
maps between these Nuprl types indexed by the arrows in {\tt C}
such that identity and composition are preserved.

We usually use letters {\tt I, J, K} $\dots$ for objects of category {\tt C}. Then if {\tt H} is a presheaf over {\tt C} then
{\tt H(I)} is a Nuprl type. We use letters $\alpha$, $\rho$ for members of a type like {\tt H(I)}  or {\tt H(J)}, but in Nuprl syntax we have to write {\tt alpha} or {\tt rho}.
An \emph{object} of preheaf {\tt H} is a pair of type {\tt I:ob(C) \mtimes{} H(I)}, so it is a pair {\tt <I, rho>} where
{\tt rho \mmember{} H(I)} (pairs in Nuprl are displayed with angle brackets). There is a natural way to define the
arrows so that the objects of a presheaf {\tt H} form a category, called the \emph{category of elements of the presheaf}.

\paragraph{Types in a context}
One way to define the meaning of {\tt H \mvdash{} T} for a presheaf {\tt H} is that {\tt T} is a presheaf over the category of elements of {\tt H}. We chose to unpack this abstract definition and
spell out what it means.

Such a {\tt T} is a functor so it has two components. The first component,  
a family of Nuprl types indexed by the objects of
{\tt H}, has type
{\tt I:ob(C) \mrightarrow{} rho:H(I) \mrightarrow{} Type}.
The second component is a family of maps between these types
indexed by the arrows of {\tt C}. When  {\tt f \mmember{} J \mrightarrow{} I} is an arrow in {\tt C} then
for {\tt rho \mmember{} H(I)} we have {\tt f(rho) \mmember{} H(J)} and this must induce a map from {\tt T(I,rho)} to
{\tt T(J,f(rho))}. So the second component of {\tt T} has type
 {\tt I:ob(C)\mrightarrow{}J:ob(C)\mrightarrow{}f:(J{}\mrightarrow{}I)\mrightarrow{}rho:H(I)\mrightarrow{}T(I,rho)\mrightarrow{}T(J, f(rho))}. Applied to {\tt I,J,f,rho} and {\tt u \mmember T(I,rho)} the second component of {\tt T} gives
a member of {\tt T(J, f(rho))} and we write this as {\tt T(I,J,f,rho,u)}.

To preserve the identity and composition we require
\begin{verbatim}
T(I,I,id(C),rho,u) = u   and 
T(I,K, f o g, rho, u) = T(J,K,g,f(rho), T(I,J,f,rho,u))
\end{verbatim}
Putting all of this together we get the definition of the
Nuprl type {\tt \{H \mvdash{} \_\}}, the type of types in context {\tt H}:
\begin{program}
\{H \mvdash{} \_\}  ==\\
\{TF:T:I:fset(\mBbbN{}) \mrightarrow{} H(I) \mrightarrow{} \mBbbU\{i\} \mtimes{}\\
    I:ob(C)\mrightarrow{}J:ob(C)\mrightarrow{}f:(J\mrightarrow{}I)\mrightarrow{}a:H(I)\mrightarrow{}T(I,a)\mrightarrow{}T(J,f(a))\\
   | let  T,F = TF  in\\
   \mforall{}I:ob(C). \mforall{}a:H(I). \mforall{}u:T(I,a). F(I,I,id,a.u) = u \mwedge{}\\
   \mforall{}I,J,K:ob(C). \mforall{}f:J\mrightarrow I.  \mforall{}g:K \mrightarrow J. \mforall{}a:H(I). \mforall{}u:T(I,a).\\
    F(I,K, f \mcdot{} g, a, u) = F(J, K, g, f(a), F(I,J,f,a ,u))\}  
\end{program}
The display form \{H \mvdash{} \_\} is meant to indicate that this
is the (Nuprl) type of things that can follow {\tt H \mvdash{}}. Then
we display {\tt T \mmember{} \{H \mvdash{} \_\}} as {\tt H \mvdash{} T}.
\paragraph{Universe levels for Contexts and Types}
In Nuprl, almost every definition is provided with a typing lemma
that we call its \emph{wellformedness lemma}. When for an expression {\tt t} and a type {\tt T} it is true the {\tt t \mmember{} T}, the {\tt Auto} tactic can usually use the wellformedness lemmas
to prove this (even though typing is undecidable in general).
Note that because of subtyping and type extensionality the type of
an expression {\tt t} is not unique, so there may be many types {\tt T} for which {\tt t \mmember{} T} is provable and {\tt Auto} may only prove some of them, while proving others may take more steps. 
The wellformedness
lemma for {\tt \{H \mvdash{} \_\}} is
\begin{program}
\mforall{}[H:\mvdash{} ]. \{H \mvdash{} \_\} \mmember{} \mBbbU\{i+1\}
\end{program} 
Although we display only {\tt \mvdash{} } for the Nuprl type of all contexts, its definition as the presheaves over
{\tt C} means that it
has a universe level parameter {\tt i}. The typing lemma says that
for any context {\tt H} (with level parameter {\tt i}), 
the type of types in context {\tt H} is a Nuprl type in universe {\tt i+1}.

Since definition of {\tt \{H \mvdash{} \_\}} mentions {\tt \mBbbU\{i\}}
the smallest universe it can be a member of is {\tt \mBbbU\{i+1\}}, and
as long as all the other types in the definition are in {\tt \mBbbU\{i+1\}}, the whole type will be in {\tt \mBbbU\{i+1\}}. So we can
allow the ``sets'' {\tt H(I)} to be in {\tt \mBbbU\{i+1\}}.
This means that we can define  {\tt \mvdash{}} to be
{\tt Presheaf(C)\{i+1\}} and still have 
{\tt \{H \mvdash{} \_\} \mmember{} \mBbbU\{i+1\}}. This subtlety turns
out to be important when we come to modeling the rules for universes
in cubical type theory.

\paragraph{Definitions of {\tt (H.T)}, {\tt p}, {\tt (T)sigma}, and empty context:}
If {\tt H} is a context and {\tt H \mvdash{} T} then {\tt H.T}
is also a context. It is a presheaf that assigns to {\tt I} the
set of pairs {\tt <rho,u>} where {\tt rho \mmember{} H(I)} and
{\tt u \mmember{} T(I,rho)}. The {\tt H}-restriction map  {\tt f(rho)}
and the ``morphism map'' {\tt T(I,J,f,rho,u)} give the {\tt H.T}-restriction map.

\begin{program}
H.T == \\
Presheaf(Set(I) = rho:H(I) \mtimes{} T(I,rho)\\
        Morphism(I,J,f,pr) = let rho,u = pr in\\
                             <f(rho), T(I,J,f,rho,u)>
        )
\end{program}
A natural transformation from {\tt H.T} to {\tt H}  must assign to each {\tt I} a map from {\tt rho:H(I) \mtimes{} T(I,rho)} to {\tt H(I)}. One
obvious choice is the polymorphic map:
\begin{program}
p ==  I |\mrightarrow{} \mlambda pr.fst(pr)
\end{program}

If {\tt H} and {\tt G} are contexts, {\tt G \mvdash{} T}, 
and {\tt sigma:H \mrightarrow{} G} then to get a type {\tt (T)sigma}
in context {\tt H} we need to define the type {\tt (T)sigma(I,rho)} and the morphism map
{\tt (T)sigma(I,J,f,rho,u)}. Since {\tt sigma} is a natural transformation, {\tt sigma(I,rho) \mmember{} G(I)} when {\tt rho \mmember{} H(I)}, so the definition is:
\begin{program}
(T)sigma == \\
<\mlambda{}I,rho. T(I, sigma(I,rho)), \\
 \mlambda{}I,J,f,rho,u. T(I,J,f,sigma(I,rho),u)>
\end{program}
For the empty context we take the presheaf 
\begin{program}
() == Presheaf(Set(I) = Unit; Morphism(I,J,f,u) = u)
\end{program}

\paragraph{Terms of type {\tt T} in context {\tt H}:}
When {\tt H \mvdash{} T}, we have ``sets'' (i.e. types) {\tt T(I,rho)} for each {\tt rho \mmember{} H(I)}. A \emph{term} {\tt t:T} will assign
to each {\tt I} and {\tt rho} a member of {\tt T(I,rho)} in a way that
respects the morphism maps. So the Nuprl type for the terms
{\tt \{H \mvdash{} \_:T\}}, i.e. the {\tt t} for which {\tt \{H \mvdash{} t:T\}}, is:

\begin{program}
\{H \mvdash{} \_:T\}  ==\\
  \{t:I:ob(C) {}\mrightarrow{} rho:H(I) \mrightarrow{} T(I,rho) |\\      
    \mforall{}I,J:ob(C). \mforall{}f:J \mrightarrow{} I. \mforall{}rho:H(I).\\
      T(I,J,f,rho,t(I,rho)) = t(J,f(rho))\}  
\end{program}
So a term in the presheaf model is a family of Nuprl terms.

Given a natural transformation {\tt sigma:H \mrightarrow{} G} and
a term {\tt t} such that {\tt G \mvdash{} t:T}, we get a term
{\tt (t)sigma} for which {\tt H \mvdash{} (t)sigma:(T)sigma}
\begin{program}
 (t)sigma == \mlambda{}I,rho. t(I,sigma(I,rho))
\end{program}

Since for the context {\tt H.T} the set {\tt (H.T)(I)} is
{\tt rho:H(I) \mtimes{} T(I,rho)}, we see that the function
\begin{program}
 q == \mlambda{}I,pr. snd(pr)
\end{program} is a term of type {\tt H.T \mvdash{} q:(T)p}.

\paragraph{Substitutions:}
We said that the \emph{context maps} {\tt sigma:H\mrightarrow{}G} 
in the name-free syntax play the role that substitutions do in a theory
with bound variables. We need one more generic definition that makes
this analogy more precise. Consider an elimination rule that says
in {\tt H, x:A \mvdash{} C} we can eliminate {\tt x} by substituting
a term {\tt u} where {\tt H \mvdash{} u:A} for {\tt x} to get
{\tt H \mvdash{} C[x/u]}. How do we express this without the bound
variable {\tt x}?

We need a context map {\tt [u]:H\mrightarrow{} H.A}, for then if
 {\tt H.A \mvdash{} C} we get {\tt H \mvdash{} C[u]}. We first make
 a more general definition:
\begin{program}
(sigma;u) ==   I |\mrightarrow{} \mlambda rho.<sigma(I,rho), u(I,rho)>
\end{program}
If {\tt sigma:H\mrightarrow{}G}, and {\tt G \mvdash{} A}, and {\tt H \mvdash{} u:(A)sigma} then for {\tt rho \mmember H(I)}, we have
{\tt sigma(I,rho) \mmember{} G(I)} and {\tt u(I,rho) \mmember{} ((A)sigma)(I,rho)}. Since {\tt ((A)sigma)(I,rho) = A(I,sigma(I,rho))}, {\tt (sigma;u) \mmember{} H\mrightarrow{} G.A}.

With the identity map {\tt 1} for {\tt sigma} we get
\begin{program}
[u] == (1,u)
\end{program}
and {\tt [u]} has type {\tt H\mrightarrow{} H.A}.

\paragraph{Basic structural rules for MLTT:}
We have given the formal Nuprl definitions for
 $(H \vdash)$, $\sigma\!:\!H \rightarrow G$, $(H \vdash T)$, $(H \vdash t:T)$, $(H.T)$, $(T)\sigma$, $(t)\sigma$, $p$, $q$, and $[u]$.
 They satisfy the following twenty basic rules.
\begin{enumerate}
\item {\tt G \mvdash{}} $\Rightarrow$   {\tt 1 \mmember{} G\mrightarrow{}G}
\item {\tt sigma \mmember{} H\mrightarrow{}G} $\wedge$ {\tt delta \mmember{} K\mrightarrow{}H} $\Rightarrow$ {\tt sigma delta \mmember{} K\mrightarrow{}G}
\item {\tt G \mvdash{} A} $\wedge$ {\tt sigma \mmember{} H\mrightarrow{}G} $\Rightarrow$ {\tt H \mvdash{} (A)sigma}
\item {\tt G \mvdash{} t:A} $\wedge$ {\tt sigma \mmember{} H\mrightarrow{}G} $\Rightarrow$ {\tt H \mvdash{} (t)sigma:(A)sigma}
\item {\tt () \mvdash{} }
\item {\tt G \mvdash{}} $\wedge$ {\tt G \mvdash{} A} $\Rightarrow$ {\tt G.A \mvdash{} }
\item {\tt G \mvdash{} A} $\Rightarrow$ {\tt  p: G.A \mrightarrow{} G }
\item {\tt G \mvdash{} A} $\Rightarrow$ {\tt G.A \mvdash{} q:(A)p }
\item {\tt sigma \mmember{} H\mrightarrow{}G} $\wedge$ {\tt G \mvdash{} A} $\wedge$ {\tt H \mvdash{} u:(A)sigma} $\Rightarrow$ {\tt (sigma,u): H\mrightarrow{}G.A }
\item {\tt 1 sigma = sigma 1 = sigma}
\item {\tt (sigma delta) nu = sigma (delta nu)}
\item {\tt [u] = (1,u)}
\item {\tt (A)1 = A}
\item {\tt ((A)sigma)delta = (A)(sigma delta)}
\item {\tt (u)1 = u}
\item {\tt ((u)sigma)delta = (u)(sigma delta)}
\item {\tt (sigma,u) delta = (sigma delta, (u)delta)}
\item {\tt p (sigma,u) = sigma}
\item {\tt (q)(sigma,u) = u}
\item {\tt (p,q) = 1}
\end{enumerate}
Notice that all of these rules are either typing rules or equations.
Thus, at least in Nuprl, there is no computational content to be extracted from the proofs of these rules. The only computational content, so far, is in the given definitions. Because of this, there
is not much reason to discuss the formal proofs of these rules, except
perhaps to note which of them depend on Nuprl's extensional features,
and just note that they have all been proved formally and are in the
Nuprl library.
 
\section{Discrete types and terms} 
We have defined a type in {\tt \{H \mvdash{} \_\}} to be a
family of Nuprl types together with a family of maps bewteen them.
One simple way to get such families is to use for the family of types
the constant family
{\tt \mlambda{}I\mlambda{}rho.A} where {\tt A} is a Nuprl type (in universe {\tt i}).
Then for the family of maps between them we can use the identity maps. Formally
this gives
\begin{program}
discrete(A) == <\mlambda{}I rho.A,\tt \mlambda{}I J f a x. x >
\end{program}
Then for any context {\tt H} we have {\tt H \mvdash{} discrete(A)}.
If {\tt t \mmember{} A} is a Nuprl term of type {\tt A} then we
have {\tt H \mvdash{} discrete(t):discrete(A)} where {\tt discrete(t)} is the constant family 
\begin{program}
discrete(t) == \mlambda{}I rho.t
\end{program}
Also, for any context map {\tt sigma: H \mrightarrow{} G} we have
{\tt (discrete(A))sigma = discrete(A)} and {\tt  (discrete(t))sigma = discrete(t)}.

When we later define the path types, we will see that the only
paths in {\tt Path(discrete(A))} will be the constant paths, {\tt refl(a)}, for {\tt a \mmember{} A}. Thus we can see the Nuprl types as types in the
cubical type theory and they are \emph{discrete} in several senses,
including the ``topological'' sense that all paths (maps from the
interval into the type) are constant.

\section{{\tt \mSigma{}} and {\tt \mPi{}} types}
\label{types} 
 
Next we define the types $\Pi(A,B)$ and $\Sigma(A,B)$ and associated terms. For a context {\tt H}, if {\tt H \mvdash{} A} and {\tt H.A \mvdash{} B}, then we should have both {\tt H \mvdash{} \mSigma(A,B)} and {\tt H \mvdash{} \mPi(A,B)}. We also need a pairing term {\tt (u,v)} to form a term {\tt w \mmember{} \mSigma(A,B)}, and 
terms  {\tt w.1} and {\tt w.2} to decompose it. We need a {\tt \mlambda}  to make {\tt f \mmember{} \mPi(A,B)} and {\tt app(f,u)} to
apply it.

Recall that to define a type {\tt T} in context {\tt H} we need
a family of types {\tt T(I,rho)} where {\tt rho \mmember{} H(I)}
and for {\tt u \mmember{} T(I,rho)} and {\tt f:J\mrightarrow{}I} we need the morphism maps {\tt T(I,J,f,rho,u) \mmember{} T(J,f(rho))}.
\paragraph{The {\tt \mSigma{}} Type:} The sigma type is relatively straightforward. Given {\tt rho \mmember{} H(I)} we have the type
{\tt A(I,rho)} and for {\tt u \mmember{} A(I,rho)} the pair {\tt <rho,u>}is a member of {\tt (H.A)(I)} so {\tt B(I,<rho,u>)} is a type.
Thus the family {\tt \mSigma(A,B)} is defined by
\begin{program}
\mSigma(A,B)(I,rho) ==  u:A(I,rho) \mtimes{} B(I,<rho;u>)
\end{program}
Given a pair {\tt p \mmember{} \mSigma(A,B)(I,rho)} and  arrow
{\tt f:J\mrightarrow{}I}, the first component of {\tt p} is a member of {\tt A(I,rho)} so {\tt A(I,J,f,rho,fst(p)) \mmember A(J,f(rho))}.
Then {\tt B(I,J,f,<rho,fst(p)>,snd(p))} is a member of {\tt B(J,f(rho,fst(p)))}. So we can define the morphism map:
\begin{program}
\mSigma(A,B)(I,J,f,rho,p) == \\ 
  <A(I,J,f,rho,fst(p)),B(I,J,f,(rho,fst(p)),snd(p))> 
\end{program}
So the formal definition of the type {\tt \mSigma(A,B)} is:
\begin{program}
\mSigma(A,B) == \\ 
 <\mlambda{} I,rho. u:A(I,rho) \mtimes{} B(I,<rho;u>),\\
  \mlambda{} I,J,f,rho,p. <A(I,J,f,rho,fst(p)),\\ 
                  B(I,J,f,<rho,fst(p)>,snd(p))>\\
 > 
\end{program}
Now if {\tt H \mvdash{} A} and {\tt H.A \mvdash{} B} we can prove {\tt H \mvdash{} \mSigma(A,B)}. Note that the definition of {\tt \mSigma(A,B)} does not mention the context {\tt H}, so it is polymorphic. That will not be the case for the {\tt \mPi}-type.

Recall that a term {\tt H \mvdash{} t:T} is a family of terms
{\tt t(I,rho) \mmember{} T(I,rho)} that respects the morphisms. 

If {\tt H \mvdash{} u:A} and {\tt H \mvdash{} v:B[u]} then the
\emph{pair term} {\tt (u,v)} is simply
\begin{program}
(u,v) == \mlambda{}I,rho. <u(I,rho),v(I,rho)>
\end{program} and for {\tt H \mvdash{} pr:\mSigma(A,B)} the terms
{\tt pr.1} and {\tt pr.2} are defined simply by:
\begin{program}
pr.1 == \mlambda{}I,rho. fst(pr(I,rho))\\
pr.2 == \mlambda{}I,rho. snd(pr(I,rho))\\
\end{program}

\paragraph{The {\tt \mPi{}} Type:} The definition of the {\tt \mPi}-type is trickier. In order to get a family of function types that will transform properly given an arrow {\tt f:J\mrightarrow{}I}, we
build that requirement into the definition. The family of types
{\tt \mPi(H,A,B,I,rho)}---now depending on the context {\tt H}--- 
is a subtype of functions of type
{\tt J:ob(C)\mrightarrow{}f:(J\mrightarrow{}I)\mrightarrow{}u:A(J,f(rho))\mrightarrow{}B(J,<f(rho),u>)
}. 
In order to transform properly, if {\tt w} is such a function, then it must satisfy the condition that for all {\tt f:J\mrightarrow{}I} and for all {\tt g:K\mrightarrow{}J}, if {\tt u\mmember A(J,f(rho))} then
\begin{program}
 B(J,K,g,<f(rho),u>,w(J,f,u)) = w(K,(f \mcdot g), A(J,K,g,f(rho),u))
\end{program}
This says applying {\tt w} and then transforming in type {\tt B} gives
the same result as first transforming in type {\tt A} and then applying {\tt w}.
Recall that restriction map {\tt f(rho)}
really has parameters {\tt H,I,J,f,rho} so this family of functions
does depend on the context {\tt H}. The formal definition is:
\begin{program}
pi-family(H;A;B;I;rho)  ==\\
\{w:J:ob(C)\mrightarrow{} f:(J\mrightarrow{}I) \mrightarrow{} u:A(J,f(rho)) \mrightarrow{} B(J,(f(rho);u)) |\\ 
 \mforall{}J,K:ob(C).\mforall{}f:J\mrightarrow{}I.\mforall{}g:K \mrightarrow{}J.\mforall{}u:A(J,f(rho)).\\
 B(J,K,g,<f(rho),u>,w(J,f,u)) = w(K,(f \mcdot g), A(J,K,g,f(rho),u))\\
\}  
\end{program}
The morphism map for this family is now easy to define. Given a {\tt w} in {\tt pi-family(H;A;B;I;rho)} and {\tt f:J\mrightarrow{}I} 
then the function 
\begin{program} \tt \mlambda K,g,v. w(K, (f \mcdot g),v)
\end{program} will be
a member of {\tt pi-family(H;A;B;J;f(rho))}. So the formal definition
of the type {\tt \mPi(A,B)} is:
\begin{program} 
\mPi{}(A,B)  == \\
   <\mlambda{}I,rho. pi-family(H;A;B;I;rho), \\   
    \mlambda{}I,J,f,rho,w,K,g. (w K (f \mcdot{} g))>
\end{program}

Now we need a {\tt \mlambda}-term to build members of {\tt \mPi{}(A,B)}. If {\tt H \mvdash{} A} and we have a term {\tt b} of type
{\tt H.A \mvdash{} b:B} (where {\tt H.A \mvdash{} B} ) then
we want to define the term {\tt \mlambda b} so that
 {\tt H \mvdash{} \mlambda b:\mPi(A,B)}.
So, given {\tt rho \mmember{} H(I)} we need a member of 
{\tt pi-family(H;A;B;I;rho)} and that is a function that takes
as input {\tt J}, {\tt f:J\mrightarrow{} I}, and {\tt u \mmember{} A(J,f(rho))}. We get such a function by applying term {\tt b} to
{\tt J} and the pair {\tt <f(rho),u>}. Thus, the definition is:
\begin{program}
(\mlambda{}b) ==  \mlambda{}I,a,J,f,u. b(J,<f(a),u>)
\end{program}
To prove the typing rule for {\tt (\mlambda{}b)} we have to show
that it is in the subtype of functions that satisfy the given
constraints. This follows from the constraints on the types {\tt A},
{\tt B}, and the term {\tt b}. 

Finally, if we have a term {\tt w} of type {\tt H \mvdash{} \mPi(A,B)}
and a term {\tt u} of type {\tt H \mvdash{} A}, then we want a term {\tt app(w,u)} of type {\tt H \mvdash B[u]}. So, given {\tt rho \mmember{} H(I)}, we have {\tt u(I,rho) \mmember{} A(I,rho)} and {\tt w(I,rho)} is a member of {\tt pi-family(H;A;B;I;rho)} so we can apply
{\tt w(I,rho)} to any {\tt J}, {\tt f:J\mrightarrow{}I} and {\tt v \mmember A(J,f(rho))}. The only sensible option is to take {\tt J = I},
{\tt f = id(I)}, and {\tt v = u(I,rho)}. The definition is therefore:
\begin{program}
app(w,u)  ==  \mlambda{}I,rho. w(I,rho)(I,1,u(I,rho))
\end{program}

\paragraph{Rules for MLTT basic type formers:}
We have given the formal Nuprl definitions for {\tt \mPi{}(A,B)}, {\tt \mSigma{}(A,B)}, {\tt app(w,u)}, {\tt (\mlambda{}b)}, {\tt (u,v)}, {\tt pr.1}, and {\tt pr.2}. 
 
 They satisfy the following nineteen basic rules.
\begin{enumerate}
\item {\tt G.A \mvdash{} B} $\Rightarrow$ {\tt G \mvdash{} \mPi(A,B)}  
\item {\tt G.A \mvdash{} B} $\wedge$ {\tt G.A \mvdash{} b:B} $\Rightarrow$ {\tt G \mvdash{} (\mlambda b):\mPi(A,B)} 
\item  {\tt G.A \mvdash{} B} $\Rightarrow$ {\tt G \mvdash{} \mSigma(A,B)}
\item {\tt G.A \mvdash{} B} $\wedge$ {\tt G \mvdash{} u:A} $\wedge$ {\tt G \mvdash{} v:B[u]} $\Rightarrow$ {\tt G \mvdash{} (u,v):\mSigma(A,B)} 
\item {\tt G \mvdash{} pr:\mSigma(A,B)} $\Rightarrow$ {\tt G \mvdash{} (pr.1):A }
\item {\tt G \mvdash{} pr:\mSigma(A,B)} $\Rightarrow$ {\tt G \mvdash{} (pr.2):B[pr.1] }
\item {\tt G \mvdash{} f:\mPi(A,B)} $\wedge$ {\tt G \mvdash{} u:A} $\Rightarrow$ {\tt G \mvdash{} app(f,u):B[u] }

\item {\tt \mPi(A,B) sigma = \mPi(A sigma, B (sigma p, q)) }
\item {\tt (\mlambda b)sigma = \mlambda(b((sigma p, q)))}
\item {\tt app(f,u)sigma = app(f sigma, u sigma)}
\item {\tt app(\mlambda b, u) = b[u]}
\item {\tt f = \mlambda(app((f)p,q)}
\item {\tt \mSigma(A,B) sigma = \mSigma(A sigma, B (sigma p, q)) }
\item {\tt (pr.1)sigma = (pr sigma).1}
\item {\tt (pr.2)sigma = (pr sigma).2}
\item {\tt (u,v)sigma = (u sigma, v sigma)}
\item {\tt (u,v).1 = u}
\item {\tt (u,v).2 = v}
\item {\tt (pr.1,pr.2) = pr}
\end{enumerate}
Again, all of these rules are either typing rules or equations
so no computational content comes from the proofs of these rules,
and the only computational content is in the given definitions.
Again, we need not discuss the proofs, all of which have been done
and are in the Nuprl library. These proofs were first done for
a particular category, the \emph{cube category}, but we later generalized these definitions and proofs to have an arbitrary
category {\tt C} as a parameter, and those are the definitions
given in this article.

The next article in the series will define the cube category
and the interval type {\tt \mBbbI}.

Full details can be found at
http://www.nuprl.org/wip/Mathematics/\hfil\break
in the two directories ``presheaf models of type theory''
and ``cubical type theory''.
The directory ``cubical sets'' contains
our (incomplete) formalization of the Bezem, Coquand, Huber model.

\bibliographystyle{plain}

\end{document}